\documentstyle[12pt,a4]{article}
\renewcommand{\baselinestretch}{1.6}
\title{Neutron activation analysis \\
of some Neolithic copper objects}
\author{Agata Olariu\\
{\em National Institute for Physisc and Nuclear Engineering}\\
{\em P.O. Box Mg-6 76900, Bucharest Magurele, Romania}}

\begin{document}
\maketitle

\section{Introduction}
In this paper we have analyzed by  neutron activation analysis 
(NAA)
a number of 93 items of Neolithic copper from National Museum of
History from Bucharest, having the  provenance from Moldavia 
region.
In the Table 1 it is shown the list of analyzed objects.

\section{Experimental method of analysis}
We have applied the NAA, the particular features of the method encountered 
for ancient copper being also detailed in reference $^{1}$.\\
{\bf Sampling}. First some corroded parts have been
removed from the surface of copper objects, the corroded material having a
totally different elemental composition from that of the body of the object. 
Then 
samples of 10-50 mg have been cut with a hard vidia knife from the object 
body and
after that washed with different solvents: acetone, benzene, ether to avoid the
impurities from the surface of the item and the protective varnish, added 
in the museum.\\   
{\bf Irradiation of medium periods}. Samples have been put in polyethylene 
foils and
irradiated at the rabbit system of the nuclear reactor VVR-S, from NIPNE 
Magurele, Bucharest at the flux of 
$\approx$1.25$x$10$^{12}$neutrons/cm$^{2}\cdot$sec, for 30 minutes.
Copper being in majority it was strongly activated so that the induced
radioactivity in the samples could be measured only after 4-5 days . 
Natural cooper has 2 isotopes:
Cu$^{63}$ and Cu$^{65}$ which by the reaction
(n, $\gamma$) give the radioisotopes 
Cu$^{64}$  T$_{1/2}$=12.74 h and Cu$^{66}$ with T$_{1/2}$=5.10 min. 
After a cooling time of 4-5 days, in the gamma spectra of the samples, the
activity coming from the photopeak of 1345.8 keV (0.0048) of 
Cu$^{64}$ is small enough and permits to remark  other elements, present
in the cooper matrix. The samples have been measured 1000 s at a spectrometric
chain using a Ge(Li) detector of 135 cm$^{3}$ and an analyzer of 4096 channels
coupled at a PC. The system gave a resolution of 2.7 keV at 1.33 keV 
(Co$^{60}$).
We observed the elements: Au, As, Cu and Sb.\\
{\bf Long time irradiation}: The samples of copper  have been wrapped in 
aluminum
foil and put it in a quartz phial together with metallic spectroscopic pure 
standards, 
copper and  nickel and irradiated at the vertical chain of the reactor, 
at a flux of 
$\approx$10$^{13}$neutrons/cm$^{2}$$\cdot$sec, for a period of time of 40 h.
After a cooling time of 2 weeks, we measured the $\gamma$ activity of the
samples at the same spectrometric chain, for 3000 s.
We have determined the following elements: Sb, Ag, Co, 
Cr, Fe, Hg, Ni, Se, Sn.\\
{\bf Cobalt}. Cobalt was determined in the copper object 
using the isotope 
Co$^{60}$ got in the reaction:
     Co$^{59}$(n, $\gamma$)Co$^{60}$. 
Co$^{60}$ is also produced by the reaction: 
 Cu$^{63}$(n, $\alpha$)Co$^{60}$ 
which is important enough in this situation, when the element
copper is the major element ($\approx$99\%).
So that  C$_{Co}$=C$_{total}$ - C$_{Co-Cu}$, where C is the concentration.
Another correction made in the calculus of the cobalt concentration in the
copper samples is that the used cooper standard contains also traces of cobalt.
It was determined that a standard of pure copper has a content of minimum 4 ppm
of cobalt.
Also in the gamma background of the experimental room it were observed the
peaks at the cobalt energies of 1773.2 keV  and 1332.5 keV; therefore from the
respective photopeaks area it was subtracted the area given by the background,
measured for the same period of time as the sample.\\
{\bf Mercury} was determined from the $\gamma$ ray of 279.2 keV and intensity
(81.5\%) of  Hg$^{203}$ with
T$_{1/2}$=46.60d. This ray is interposed with the $\gamma$ ray from
Se$^{75}$, of  279.5 
keV, and intensity 0.25. Therefore the contribution of the mercury must be
extracted from the peak of peak of 279 keV: 
N$_{Hg^{203}}$=N$_{total 279keV}$ - N$_{Se^{75}}$, 
where N$_{Hg^{203}}$ is the counting rate in the peak of 279 keV, 
given by mercury contribution.
N$_{total 279keV}$ is the total counting rate 
in the peak of 279 keV\\
N$_{Se^{75}}$ is the counting rate in the peak of 279 keV, due
to the selenium presence.\\
It was used as reference the selenium peak from the energy 
264.7 keV of  intensity of 0.5658 :\\
$\epsilon_{264keV}\cdot$ s$_{264keVSe^{75}}\cdot$ N$_{264keVSe^{75}}$=
$\epsilon_{279keV}\cdot$ s$_{279keVSe^{75}}\cdot$ N$_{279keVSe^{75}}$
where:
$\epsilon_{264keV}$ is the efficiency of the  detector from 264 keV, 
$\epsilon_{279keV}$, is the efficiency  at the  energy of 
279 keV, 
s$_{264keVSe^{75}}$, the intensity of the  line of 264 keV of Se$^{75}$, 
s$_{279keVSe^{75}}$, the intensity of the line of 279 keV of Se$^{75}$\\
{\bf Nickel}. The concentration of nickel was measured by the isotope 
Co$^{58}$ (T$_{1/2}$=71.3 d). Nickel was a exception by the fact that it was 
determined by the reaction 
Ni$^{58}$(n, p)Co$^{58}$, unlike the other elements determined by the reaction
(n, $\gamma$). 

\section{Results of analysis}

In the Table 2 are given the results of activation analysis for the 
Neolithic  copper
objects, from the National Museum of History from Bucharest.
The concentrations are given in ppm, and when an element was determined 
in a quantity larger than 10000 ppm, its concentrations was expressed 
in percents, using the notation of \%. The measured errors were the statistical
errors and were in mean of $<$10\%. 
In the situations when the signal was dimmed by the background $\gamma$ Compton
the result was given as -, with the significance of {\it under the limit of 
detection.} \\
The NAA provides the values of the concentrations for the determined
elements in the Neolithic copper objects that establish a basis for further
characterizations and interpretations together with the considerartion of 
historic data as culture, dating or style.

{\bf Acknowledgement}\\
We thank archaeologist Dragomir Popovici for collaboration.\\

\newpage

{\bf \large References}\\
1. Agata Olariu, C. Besliu, M. Belc, I. V. Popescu, T. Badica,
Compositional Studies of Ancient Copper from Romanian
Territories,
Los Alamos e-print Archive, nucl-ex, paper 9907015, and
Journal of Radioanalytical and Nuclear Chemistry, 1999
\newpage
\renewcommand{\baselinestretch}{1.5}
\setlength{\textwidth}{16.5cm}
\pagestyle{empty}
\small
\begin{table}
\newlabel{}
\caption{{\bf Table 1}. List of analyzed Neolithic copper objects,
National Museum of History Bucharest}\\

\begin{tabular}{lccc}
\hline
Sample  & Object of copper & Reg. no. & Hoard, provenance\\
\hline
\hline
P1 & Axe-Pick-axe     & 170 & Central Military Museum\\
P2 & Axe              & 169 & Central Military Museum \\
P3 & Axe              & 170 & Central Military Museum\\
P4 & Axe              & 36241 & Central Military Museum\\
P5& Copper object     & 6     & Vaslui Museum, Fedesti Cetate\\
P6& Needle            & 13444 & Vaslui Museum, Dumesti\\
P7 & Needle           &       & Malnas, Cucuteni A\\
P8& Axe of copper     &       & Fastici, Vaslui county (Import?)\\
P9& Needle (?)        &   94  &  Malnas\\
P10& Bead frag.       &       & Brad hoard, Bran Museum\\
P11& Bracket          & 17579 & Brad hoard\\
P12& Bracket          &       & Brad hoard\\
P13& Bracket          & 17578 & Brad hoard\\
P14& Axe              &       & Brad hoard\\
P15& Bracelet         & 17577 & Brad hoard\\
P16& Bracelet         & 17576 & Brad hoard\\
P17&Rite (?) axe      &17575  & Brad hoard      \\
P18&Axe-Pick-axe      & 740   & Slobozia-Bodoganesti, Museum Complex Iasi\\
P19& Axe              &       & Erbiceni,  Iasi county\\
P20& Chisel           &11145  & Rus/81, Rusaesti-Poduri, Piatra Neamt\\
P21& Needle           & 11143 & Rus/81, Rusaesti-Poduri\\
P22& Needle (?)       & 11143 & Rusaesti-Poduri\\
P23& Metallic frag.   & 11142 & Rusaesti-Poduri\\ 
P24& Needle           & 11146 &Rusaesti-Poduri\\
P25& Needle           & 11148 &Rusaesti-Poduri\\
P26& Metallic sheet   & 11141 &Rusaesti-Poduri\\
P27& Needle           & 11147 &Rusaesti-Poduri\\
P28& Needle of copper & 11144 & Rusaesti-Poduri\\
P29& Frag. of bead    & II 14910-6395 &Traian, 1953, Neamt county\\
P30& Needle           & 6443  & Traian, 1957\\

\end{tabular}
\end{table}

\begin{table}
\begin{tabular}{lccc}
\hline
Sample  & Object of copper & Reg. no. & Hoard, provenance\\
\hline
\hline
P31&Needle     & 894 & Izvoare 1939, Neamt county\\
P32&Needle     & 894 & Izvoare 1939\\
P33&Needle     & 894 & Izvoare 1939\\
P34&Needle     & 894 & Izvoare 1939\\
P35&Piece of copper& 1615 & Izvoare\\
P36&Metallic frag. & 1544& Podei, Tg. Ocna\\
P37&Small bead & 6394 & Traian, 1952\\
P38&Needle &1543& Podei, Tg. Ocna 1943\\
P39& Needle & 6440 & Traian, 1950\\
P40&Needle & 6440& Traian, 1957\\
P41&Needle & 6438 & Traian 1956\\
P42&Needle & 6439 & Traian, 1952\\
P43&Needle & 6445 & Traian 1957\\
P44&Needle of ornament& 6370 & Tarpesti, Neamt county 1963\\
P45&Spiral needle for hair & 6374& Tarpesti 1963\\
P46& Wire & 6608 & Tarpesti 1962\\
P47& Dagger  & 1330 & Frumusica, Neamt county\\
P48& Link & 15526& Rusaesti/86, cassette A\\
P49& Needle  &   6609 & Tarpesti, 1963\\
P50& Needle  &   6615 & Tarpesti, 1965\\
P51& Frag. of copper    & 6617  & Tarpesti, 1963\\
P52& Frag. od needle & 6590 & Tarpesti, 1962\\
P53& Disk  &6593& Tarpesti L11\\
P54& Frag. of bracelet & 6595 & Tarpesti, 1964\\
P55& Frag. of bronze  & 6597 & Tarpesti, 1959\\
P56& Frag. of needle   & 6596 & Tarpesti, 1963\\
P57&Frag. angling rod (?)&6599& Tarpesti, 1962\\
P58&Peack  & 6600 & Tarpesti, 1962\\
P59&Needle frag. &6601 & Tarpesti, 1962 L7\\
P60& Miniature Axe &6591 & Tarpesti, 1962\\
P61& Brass Needle  & 6613 & Tarpesti, 1968\\
P62& Wire & 7923 & Tarpesti, 1968\\
P63&Small brass hook & 6587 & Tarpesti, 1968\\
P64& Sheet   & 6594    &Tarpesti, 1963 \\

\end{tabular}
\end{table}

\begin{table}
\begin{tabular}{lccc}
\hline
Sample  & Object of copper &Reg. no. & Hoard, provenance\\
\hline
\hline
P65&Needle     & 6436    & Traian, 1954\\
P66&Frag. bracelet & 15527  & Poduri-Rusaesti 86\\
P67&Needle & 6437   &    Traian, 1952\\
P68&Needle & 6444   & Traian, 1958\\
P69& Rolled Sheet  & 6453 & Traian, 1952\\
P70&Needle for angling rod &  6407 & Traian, 1957\\
P71&Spiral   & 6452 & Traian, 1952\\
P72&Metallic frag.  & 894 & Izvoare, 1939\\
P73&Wedding ring    & 6403 & Traian, 1958\\
P74&Needle            & 15528 & Izvoare, 1984 L10\\
P75&Needle            & 894   & Izvoarele \\
P76&Idol "en violon"& 6451 & Traian, 1952\\
P77& Spiral bracelet & 779   &Izvoare\\ 
P78&Needle(?)&   & Piatra Neamt\\
P79&Needle(?)&   & Piatra Neamt\\
P80&Needle(?)&   & Piatra Neam't\\
P81& Axe &3292 & Dragomiresti\\
P82& Axe passim & 4697& Sarata, Piatra Neamt\\
P83& Axe        & 4696       & Viisoara, Manastirea Bistritei\\
P84&Needle& 5594&Calu 1974, Neamt county\\
P85&Frag. metalic& 2958 & Podei, 1956\\
P86& Axe & 965& Floresti, Vaslui county \\
   &      &    & Vasile Parvan Museum, Barlad\\        
P87& Axe &964  & Floresti, Vaslui county\\
P88& Axe & 7034& Lupesti, Vaslui county\\
P89& Axe & 7984       & Falciu, novelty, import(?)\\
P90& Axe & 975        & Bacesti, Vaslui county\\
P91&Needle of copper & 8016 & Trestiana, Grivita, Vaslui county \\
P92&Small bead  &8543 &  Falciu, the same complex to P89\\
P93&Small bead &8543 &  Falciu\\
\hline
\end{tabular}
\end{table}
\begin{table}
\newlabel{}
\caption{{\bf Table 2}. Concentrations of analyzed Neolithic copper objects, by 
NAA}\\

\begin{tabular}{lcccccccccccc}
\hline
Sample  & Au   & As &     Sb   & Se        & Hg     & Cr
& Ag  & Ni & Fe & Zn & Co & Sn\\
\hline
\hline
                
P1	&	0.1	&	--	&	2	&	13	&	--	&	133	&	14.6	&	130	&	7680	&	426	&	3.2	&	--	\\
P2	&	0.4	&	--	&	3	&	37	&	2	&	219	&	11	&	160	&	1.01\%	&	708	&	0.9	&	--	\\
P3	&	0.2	&	14	&	2	&	162	&	3.4	&	160	&	17	&	430	&	8200	&	452	&	29	&	--	\\
P4	&	0.3	&	--	&	18	&	32	&	2.8	&	110	&	23.5	&	150	&	4960	&	257	&	1.6	&	--	\\
P5	&	0.3	&	--	&	1	&	56	&	0.3	&	40	&	11.5	&	178	&	298	&	6	&	4.1	&	--	\\
P6	&	24.3	&	--	&	3	&	0.2	&	1.2	&	29	&	358	&	197	&	--	&	6.9	&	5.3	&	230	\\
P7	&	1.4	&	7.6	&	100	&	60	&	1.4	&	--	&	54	&	97	&	--	&	6	&	0.6	&	--	\\
P8	&	0.2	&	--	&	1	&	551	&	84	&	196	&	50	&	250	&	9660	&	524	&	3.5	&	--	\\
P9	&	$<$0.1  &	29.6	&	1	&	--	&	14	&	70	&	6	&	--	&	1.49\%	&	69	&	1	&	--	\\
P10.1	&	13	&	--	&	11	&	309	&	1.8	&	--	&	518	&	206	&	--	&	--	&	1.1	&	--	\\
P10.2	&	5.1	&	9.2	&	27	&	856	&	3.4	&	236	&	75	&	201	&	1080	&	510	&	5.4	&	--	\\
P10.3	&	6.4	&	10.7	&	4.3	&	300	&	1.6	&	110	&	885	&	200	&	1580	&	187	&	1.2	&	--	\\
P10.4	&	--	&	4.3	&	11	&	329	&	1.7	&	797	&	418	&	150	&	900	&	51	&	0.7	&	--	\\
P11	&	15.2	&	1380	&	2370	&	1010	&	$<$25	&	--	&	828	&	2400	&	--	&	--	&	110	&	3.82\%	\\
P12	&	2.2	&	5303	&	835	&	120	&	--	&	--	&	182	&	1.72\%	&	--	&	--	&	90	&	13.2\%	\\
P13	&	80	&	173	&	92	&	35	&	2.5	&	153	&	98	&	870	&	7360	&	270	&	1.2	&	330	\\
P14	&	1	&	24	&	32	&	24	&	2	&	--	&	218	&	134	&	--	&	16	&	1.4	&	--	\\
P15	&	2.6	&	--	&	5	&	370	&	10	&	270	&	885	&	240	&	1.24\%	&	661	&	4	&	400	\\
P16	&	5.5	&	--	&	5	&	395	&	--	&	150	&	914	&	190	&	--	&	561	&	3	&	--	\\
P17	&	2.9	&	--	&	0.4	&	4385	&	--	&	360	&	46	&	278	&	7520	&	504	&	1.5	&	620	\\
P18	&	--	&	--	&	$<$1	&	--	&	1.5	&	--	&	6.4	&	20	&	--	&	17	&	12	&	--	\\
P19	&	5.1	&	265	&	260	&	260	&	3.4	&	70	&	370	&	760	&	1.13\%	&	330	&	2.1	&	--	\\
P20	&	0.7	&	--	&	3	&	180	&	0.1	&	--	&	22	&	195	&	--	&	25	&	3.6	&	--	\\
P21	&	--	&	--	&	1	&	157	&	0.7	&	--	&	22	&	654	&	1.38\%	&	460	&	130	&	--	\\
P22	&	$<$0.2	&	13.8	&	1	&	140	&	2.3	&	--	&	22	&	512	&	1.19\%	&	404	&	110	&	--	\\
P23	&	--	&	--	&	1.3	&	177	&	3	&	--	&	18.5	&	936	&	2.65\%	&	560	&	150	&	--	\\
P24	&	4	&	35.4	&	49	&	163	&	--	&	--	&	839	&	380	&	1500	&	180	&	4	&	--	\\
P25	&	1.6	&	8.2	&	5	&	58	&	11	&	485	&	150	&	130	&	1.91\%	&	1500	&	1	&	--	\\
P26	&	7	&	--	&	41	&	636	&	3.1	&	--	&	210	&	85	&	--	&	20	&	1	&	--	\\
P27	&	0.1	&	--	&	5	&	24	&	--	&	--	&	39	&	90	&	--	&	20	&	1	&	--	\\
P28	&	2	&	--	&	17	&	210	&	--	&	--	&	100	&	180	&	--	&	10	&	2	&	--	\\
P29	&	5.8	&	--	&	1	&	448	&	--	&	190	&	117	&	60	&	--	&	--	&	2	&	--	\\
P30	&	14	&	$<$10	&	4	&	96	&	3.3	&	--	&	34	&	120	&	--	&	10	&	3	&	--	\\

\end{tabular}
\end{table}
\begin{table}
\begin{tabular}{lcccccccccccc}
\hline
Sample  & Au & As & Sb & Se & Hg & Cr & Ag & Ni & Fe & Zn & Co & Sn\\
\hline
\hline

P31	&	--	&	--	&	39	&	774	&	0.9	&	--	&	120	&	50	&	--	&	12	&	2	&	--	\\
P32	&	--	&	1990	&	910	&	90	&	--	&	--	&	870	&	890	&	--	&	--	&	20	&	9.57\%	\\
P33	&	--	&	--	&	13	&	235	&	6.3	&	240	&	354	&	185	&	--	&	$<$20	&	6	&	--	\\
P34	&	2.2	&	7	&	52	&	143	&	0.7	&	--	&	500	&	160	&	--	&	20	&	10	&	--	\\
P35	&	92	&	1025	&	880	&	130	&	$<$10	&	4220	&	765	&	360	&	1.84\%	&	3.08\%	&	34	&	2.5\%	\\
P36	&	--	&	1.7\%	&	35	&	100	&	0.7	&	$<$23	&	79	&	60	&	--	&	10	&	2	&	--	\\
P37	&	--	&	4150	&	750	&	1740	&	8	&	$<$160	&	580	&	--	&	1.27\%	&	744	&	4	&	--	\\
P38	&	4	&	1.25\%	&	45	&	890	&	3	&	$<$70	&	115	&	110	&	--	&	--	&	2	&	--	\\
P39	&	64	&	$<$10	&	4	&	6	&	--	&	170	&	27	&	74	&	9040	&	491	&	4	&	260	\\
P40	&	318	&	$<$20	&	1	&	29	&	3	&	190	&	33	&	100	&	1.15\%	&	643	&	4	&	340	\\
P41	&	71	&	5320	&	507	&	315	&	9.7	&	--	&	1115	&	2510	&	1.69\%	&	947	&	13	&	3.43\%	\\
P42	&	$<$10	&	--	&	0.9	&	--	&	$<$3	&	275	&	5	&	100	&	1.55\%	&	880	&	5	&	--	\\
P43	&	$<$13	&	80	&	32	&	120	&	3	&	715	&	75	&	690	&	4.396\%	&	2410	&	10	&	1240	\\
P44	&	--	&	3590	&	2890	&	130	&	--	&	--	&	94	&	5690	&	--	&	--	&	80	&	28.5\%	\\
P45	&	$<$40	&	290	&	68	&	125	&	16	&	860	&	480	&	550	&	2.89\%	&	1846	&	20	&	2.52\%	\\
P46	&	--	&	$<$5	&	3.5	&	24	&	0.2	&	--	&	15	&	160	&	--	&	10	&	2	&	260	\\
P47	&	$<$50	&	820	&	100	&	300	&	30	&	--	&	5100	&	430	&	--	&	$<$100	&traces	&	--	\\
P48	&	--	&	20	&	2.8	&	3	&	--	&	--	&	8	&	170	&	--	&	traces	&	2	&	--	\\
P49	&	0.3	&	$<$5	&	0.3	&	1040	&	0.7	&	352	&	19	&	440	&	$<$200	&	10	&	5	&	--	\\
P50	&	1.2	&	13	&	7.5	&	10	&	$<$2	&	--	&	23	&	230	&	$<$200	&	75	&	2	&	460	\\
P51	&	1	&	13	&	2	&	265	&	2.4	&	--	&	32	&	270	&	--	&	40	&	12	&	--	\\
P52	&	99	&	3485	&	830	&	2910	&	--	&	--	&	1590	&	--	&	--	&	--	&	1	&	--	\\
P53	&	36	&	52	&	853	&	160	&	9.4	&	--	&	3250	&	390	&	--	&	--	&	1	&	--	\\
P54	&	12.5	&	4350	&	8517	&	--	&	--	&	--	&	42	&	3140	&	9970	&	490	&	70	&	23.8\%	\\
P55	&	32.5	&	1900	&	430	&	110	&	--	&	--	&	524	&	780	&	--	&	5.9\%	&	20	&	7.56\%	\\
P56	&	25	&	170	&	32	&	4710	&	--	&	1200	&	2239	&	549	&	6.92\%	&	4086	&	16	&	--	\\
P57	&	23	&	$<$19	&	10.3	&	334	&	2	&	--	&	1544	&	340	&	--	&	40	&	1	&	--	\\
P58	&	21	&	7150	&	1.22\%	&	--	&	--	&	--	&	3900	&	5327	&	--	&	--	&	30	&	7.07\%	\\
P59	&	2	&	78	&	76	&	--	&	--	&	1410	&	20	&	$<$470	&	5.79\%	&	4\%	&	14	&	--	\\
P60	&	15	&	--	&	5	&	190	&	--	&	$<$100	&	832	&	135	&	4220	&	250	&	2	&	--	\\
P61	&	17.5	&	4370	&	2440	&	40	&	--	&	--	&	250	&	5170	&	--	&	--	&	220	&	28.02\%	\\
P62	&	9	&	1100	&	1044	&	80	&	$<$7	&	--	&	600	&	1690	&	--	&	50	&	81	&	21.82\%	\\
P63	&	28.1	&	297	&	490	&	140	&	--	&	--	&	1320	&	1220	&	--	&	--	&	1	&	500	\\
P64	&	0.6	&	20	&	16	&	8	&	$<$3	&	--	&	73	&	120	&	--	&	23	&	2	&	--	\\

\end{tabular}
\end{table}

\begin{table}
\begin{tabular}{lcccccccccccc}
\hline
Sample  & Au & As & Sb & Se & Hg & Cr & Ag & Ni & Fe & Zn & Co & Sn\\
\hline
\hline

P65	&	150	&	--	&	6	&	60	&	17	&	500	&	45	&	224	&	1.65\%	&	1270	&	5	&	--	\\
P66	&	$<$0.6	&	--	&	1	&	20	&	$<$4	&	--	&	20	&	90	&	--	&	10	&	1	&	--	\\
P67	&	2	&	--	&	2	&	30	&	8	&	345	&	9	&	$<$100	&	2.04\%	&	1160	&	6	&	--	\\
P68	&	$<$0.2	&	10	&	3	&	38	&	1	&	128	&	5	&	65	&	6080	&	350	&	5	&	--	\\
P69	&	11.2	&	12	&	23	&	495	&	3.2	&	--	&	542	&	215	&	--	&	--	&	1	&	--	\\
P70	&	8.8	&	$<$7	&	3	&	45	&	0.8	&	620	&	21	&	180	&	7640	&	596	&	5	&	--	\\
P71	&	8.5	&	$<$5	&	1.5	&	350	&	--	&	130	&	210	&	115	&	7440	&	387	&	7	&	--	\\
P72	&	traces	&	5	&	1	&	518	&	0.4	&	--	&	19	&	275	&	--	&	7	&	4	&	--	\\
P73	&	1.6	&	660	&	137	&	310	&	5.2	&	$<$40	&	110	&	810	&	5100	&	340	&	40	&	2.89\%	\\
P74	&	5	&	$<$5	&	12	&	178	&	4	&	77	&	247	&	170	&	4100	&	230	&	2	&	--	\\
P75	&	3.9	&	--	&	3	&	190	&	5	&	--	&	355	&	220	&	--	&	--	&	1	&	--	\\
P76	&	25	&	--	&	125	&	590	&	255	&	--	&	463	&	180	&	--	&	--	&	1	&	--	\\
P77	&	5.5	&	$<$5	&	1.5	&	107	&	0.6	&	--	&	11	&	150	&	--	&	6	&	1	&	--	\\
P78	&	1.6	&	$<$6	&	7	&	42	&	--	&	$<$40	&	25	&	230	&	$<$200	&	17	&	2	&	--	\\
P79	&	38.7	&	1420	&	357	&	85	&	3.3	&	--	&	310	&	3950	&	5080	&	310	&	75	&	27.83\%	\\
P80	&	11.2	&	8	&	32	&	87	&	5	&	150	&	230	&	245	&	7650	&	456	&	4	&	920	\\
P81	&	13.2	&	--	&	31	&	7.37\%	&	--	&	$<$100	&	459	&	118	&	3400	&	190	&	2	&	--	\\
P82	&	$<$0.2	&	13	&	5	&	804	&	1.3	&	72	&	10	&	201	&	2720	&	154	&	2	&	110	\\
P83	&	1.1	&	$<$4	&	8	&	37	&	0.8	&	80	&	75	&	116	&	2840	&	143	&	2	&	--	\\
P84	&	0.2	&	6	&	2	&	3	&	$<$4	&	70	&	15	&	80	&	--	&	16	&	3	&	--	\\
P85	&	4.7	&	250	&	676	&	210	&	16	&	--	&	340	&	--	&	4140	&	1.18\%	&	4	&	57.91\%	\\
P86	&	--	&	46	&	4	&	10	&	$<$2	&	177	&	12	&	40	&	5850	&	460	&	7	&	--	\\
P87	&	0.4	&	--	&	4	&	18	&	--	&	70	&	18	&	150	&	3750	&	227	&	2	&	--	\\
P88	&	0.7	&	15	&	5	&	6	&	$<$1	&	30	&	21	&	170	&	2610	&	130	&	2	&	--	\\
P89	&	1.8	&	27	&	1160	&	40	&	--	&	--	&	74	&	340	&	4800	&	250	&	2	&	--	\\
P90	&	83	&	8240	&	17	&	478	&	0.54	&	215	&	265	&	210	&	1.025\%	&	580	&	4	&	--	\\
P91	&	7.7	&	1600	&	320	&	425	&	0.03	&	200	&	210	&	440	&	1\%	&	520	&	2	&	--	\\
P92	&	0.7	&	3	&	1	&	4	&	1.8	&	20	&	5	&	162	&	1550	&	55	&	2	&	--	\\
P93	&	1.1	&	$<$3	&	1	&	3	&	2	&	30	&	5	&	160	&	1500	&	50	&	2	&	--	\\
\hline
-=under &the limit& of det.&       & & &&&&&&&\\
\hline
\end{tabular}
\end{table}
\end{document}